\documentclass[conference]{IEEEtran}

\usepackage{graphicx}
\usepackage{subfigure}
\usepackage{amsmath}
\usepackage{amsfonts}
\usepackage{amssymb}
\usepackage{latexsym}
\usepackage{moreverb}
\usepackage{booktabs}
\usepackage{times}
\usepackage{indentfirst}
\usepackage{algorithmic}
\usepackage{algorithm}
\usepackage[table]{xcolor}  
\usepackage{multirow}       
\usepackage{array}
\usepackage{picins}
\usepackage{wrapfig}
\usepackage{fancyhdr}
\graphicspath{{figures/}}

\newcommand{\Fig}[1]{Fig.~\ref{#1}}

\newcommand{\Sec}[1]{Section~\ref{#1}}

\newtheorem{dfn}{Definition}

\fancypagestyle{IEEE_Green_open_access_footer}{%
  \fancyhf{}
  
  \fancyfoot[C]{\footnotesize \copyright2017 IEEE. Personal use of this material is permitted. Permission from IEEE must be obtained for all other uses, in any current or future media, including reprinting/republishing this material for advertising or promotional purposes, creating new collective works, for resale or redistribution to servers or lists, or reuse of any copyrighted component of this work in other works. DOI: 10.1109/ICC.2017.7997193}
}

\begin{document}


\title{Are You in the Line? \\ RSSI-based Queue Detection in Crowds}

\author{Fang-Jing~Wu and G\"urkan~Solmaz \\
NEC Laboratories Europe, Heidelberg, Germany\\
fang-jing.wu@neclab.eu, gurkan.solmaz@neclab.eu
}

\maketitle
\thispagestyle{IEEE_Green_open_access_footer}

\begin{abstract}
Crowd behaviour analytics focuses on behavioural characteristics of groups of people instead of individuals' activities. This work considers \emph{human queuing behaviour} which is a specific crowd behavior of groups. We design a plug-and-play system solution to the \emph{queue detection problem} based on Wi-Fi/Bluetooth Low Energy (BLE) received signal strength indicators (RSSIs) captured by multiple signal sniffers. The goal of this work is to determine if a device is in the queue based on \emph{only} RSSIs. The key idea is to extract features not only from individual device's data but also mobility similarity between data from multiple devices and mobility correlation observed by multiple sniffers. Thus, we propose \emph{single-device feature extraction}, \emph{cross-device feature extraction}, and \emph{cross-sniffer feature extraction} for model training and classification. We systematically conduct experiments with simulated queue movements to study the detection accuracy. Finally, we compare our signal-based approach against camera-based face detection approach in a real-world social event with a real human queue. The experimental results indicate that our approach can reach minimum accuracy of $77\%$ and it significantly outperforms the camera-based face detection because people block each other's visibility whereas wireless signals can be detected without blocking.

\end{abstract}


\section{Introduction}

Recently, \emph{crowd behaviour analytics} has attracted much attention and has boosted many promising applications such as crowd detection and estimation for public safety \cite{Li2015_Wi-Counter}\cite{2015Li_SenseFlow}, social activity analytics \cite{Santani_2016:SocailNightActivities}, and space syntax analytics for exploring new business opportunities \cite{Mashhadi2016-Wi-FiAnalytics}. These applications exploit Internet-of-Things (IoT) sensing technology with ambient sensors, Wi-Fi and Bluetooth Low Energy (BLE) sniffers, and built-in sensors of smartphones to capture human behaviour. Meanwhile, many research efforts have paid attention to individual human activity analytics such as transportation activity detection \cite{wu2014_UrbanMobilitySense} and daily activity recognition \cite{Riboni2016-DailyActivityRecognition}. However, compared to the human activity detection, crowd behaviour analytics focuses more on the behavioural characteristics of targeted groups of people instead of individual activities.

\begin{figure}
\centering
\includegraphics[width=\columnwidth]{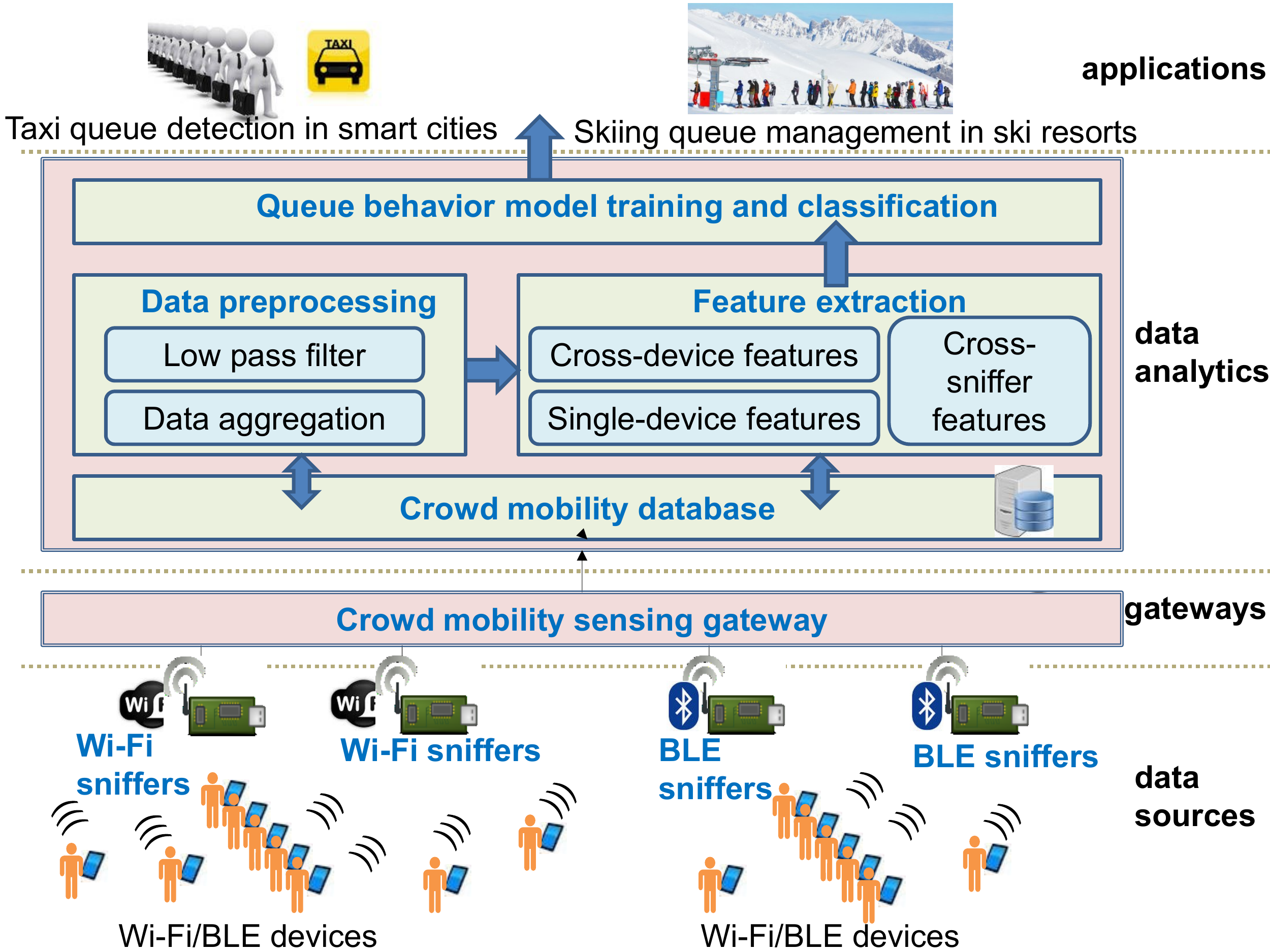}
\caption{An overview of queuing behaviour detection system.} \label{Fig:SystemOverview}
\end{figure}

This work focuses on \emph{human queuing behaviour} ("queuing behaviour" for short) which is a unique type of crowd behaviour created by groups of people rather than each individual. In reality, humans queue up for a specific purpose such as buying tickets, taking lifts in ski resorts, and waiting for taxis in which the starting point of a queue is known. Their movements observed by other people have unique patterns such as movements towards the same direction, small-scale and slow movements, and periodic movements. On the other hand, since Wi-Fi/BLE-enabled smart devices (e.g., smartphones and wearable devices) become more popular, the RSSIs in wireless packets from these devices carried by crowds provide insightful clues to capture these unique crowd mobility patterns. Thus, we consider wireless signal sniffers as crowd mobility observers which will capture broadcast packets (e.g, Wi-Fi probe request packets and BLE advertising packets) from mobile devices. This paper considers the BLE technology to implement a proof-of-concept prototype due to the privacy concern, where opt-in data collection is adopted \footnote{For experimental purpose, we collected data only from a specific set of BLE beacons carried by participants.}. The system consists of multiple signal sniffers to capture queuing behaviour in crowds and extract feature patterns of queuing behaviour based on \emph{only} RSSIs. \Fig{Fig:SystemOverview} shows an overview of our system which contains data sources, gateways, data analytics, and applications. These layers and the system is explained in detail in Section~\ref{Sec:ProblemStatementAndSystemOverview}. In this work, we deploy three sniffers to observe RSSI changes of crowds' devices \footnote{In general, more sniffers can provide richer information for detecting queuing behaviour, and we can deploy these sniffers strategically.}. One sniffer is deployed at the starting point of a potential queue such as a service counter, and another two are deployed at the left-hand side and the right-and side along the queue. The goal of our system is to determine if a person is in the queue based on the changes of RSSIs.

However, our system uses \emph{only} RSSI information to extract queueing behaviour patterns which raises the following technical challenges.
\begin{itemize}
  \item \emph{Noisy and fluctuating RSSIs}: The RSSIs may vary dramatically over time, even in the case that people are staying static and closer to the sniffers.
  \item \emph{Heterogeneity of antenna sensitivity}: The sensitivity of devices' antenna are different from each other. Specifically, the RSSIs of same device observed by different sniffers may be very different from each other. Similarly, when multiple devices carried by the same person, the RSSIs of these devices observed by the same sniffer have different variations.
  \item \emph{Cross-entity feature extraction}: Since queuing behaviour is created by groups of people, extracting feature patterns between multiple devices based on multiple observers (e.g., sniffers) is a new challenge.
\end{itemize}

To address the above technical issues, the key idea of our work is to extract feature patterns of crowd behaviour not only from each individual device's data but also from cross-device data and cross-sniffer data. Specifically, we consider three types of feature extraction: (1) \emph{single-device feature extraction}, (2) \emph{cross-device feature extraction}, and (3) \emph{cross-sniffer feature extraction}. The first type of features considers each individual device's RSSI variation when the device's owner makes movements along the queue. The second type of features considers the mobility similarity between multiple devices when these devices' owners make movement together along the queue. The third type of features considers the mobility correlation observed by multiple sniffers when a particular device's owner makes movements along the queue. Based on the three types of features extracted from individuals and heterogeneous entities, we use well-known classifiers implemented by Weka \cite{Weka} to verify the merit of these features. Note that our work mainly focuses on designing a holistic and plug-and-play solution to the queue detection problem using Internet-of-Things (IoT) sensing technology instead of designing classification algorithms. The proposed proof-of-concept prototype system can be extended to the Wi-Fi-based sniffing technology, where Wi-Fi probe request packets are captured for queue detection. We systematically conduct experiments with simulated queue movements to investigate how different parameters and setups affect detection accuracy. Finally, we compare our signal-based approach against camera-based face detection approach in a real-world team-building social event, where people queued up for taking cakes, biscuits, and drinks. The experimental results indicate that our approach can reach minimum accuracy of $77\%$ and it significantly outperforms the camera-based face detection approach since in the RSSI-based detection wireless signals are not blocked by the crowd themselves as in the case of camera-based detection.


\section{Related Work}

Recently, some research studies have paid attention to queue detection \cite{Barbagli2011_TrafficQueue} \cite{Lu2015_Taxi-PassengerQueue}\cite{Zanin2003_image-BasedVehicleQueue}\cite{Satzoda2012_Vision-basedVehicleQueue}\cite{Li2014_QueueSense}\cite{Li2016_QueueSenseTMC}\cite{Wang2014_TrackingHumanQueues}. In \cite{Barbagli2011_TrafficQueue}, an acoustic sensor network is deployed along road segments to monitor traffic queue. The work in \cite{Lu2015_Taxi-PassengerQueue} designs a system to detect queuing activities of taxis and passengers based on their GPS information. Research efforts in \cite{Zanin2003_image-BasedVehicleQueue}\cite{Satzoda2012_Vision-basedVehicleQueue} consider video-based approaches to detect vehicle queues. In \cite{Li2016_QueueSenseTMC}, a mobile application is designed to capture human behaviour and detect queueing behaviour, where built-in sensors (e.g., accelerometers, compasses, and Bluetooth) are exploited for capturing human mobility information. The work in \cite{Wang2014_TrackingHumanQueues} proposes an RSSI-based approach to detect three statuses of queuing behaviour including waiting period, service period, and leaving period based on a single Wi-Fi sniffer's observations.

However, sensor-based approaches require well-deployed infrastructure, while GPS-based approaches work only in outdoor environments. Mobile sensing approaches require mobile applications to be pre-installed in people's smartphones. Video-based approaches may compromise personal privacy and they are easily affected by light conditions and crowd blocking of visible area. The RSSI-based approach in \cite{Wang2014_TrackingHumanQueues} focuses on RSSI features from each individual device, while we consider not only single-device features but also cross-device features and cross-sniffer features since unique behavioural characteristics of human queues are created by groups of people as opposed to individuals.

\section{Queuing Behaviour Detection}\label{Sec:ProblemStatementAndSystemOverview}

\subsection{Queue Detection Problem}
Assume that the starting point of a queue (e.g., service counter) is known. We consider three signal sniffers, denoted $\pi_1$, $\pi_2$, and $\pi_3$, where the BLE sniffing technology is used in this work due to privacy concern. One sniffer is deployed in the starting point and another two sniffers are symmetrically deployed along a straight line from the stating point towards the direction of a human queue. Suppose that each visitor carries a BLE transmitter in a designated environment (e.g., exhibitions). For a given time window $\Delta_k=[t_i, t_j)$, there are three time series data streams captured by the deployed three sniffers, denoted by $\Omega_1(\Delta_k)$, $\Omega_2(\Delta_k)$, and $\Omega_3(\Delta_k)$, where each $\Omega_i(\Delta_k)$ is a sequence of BLE advertising packets. Each BLE advertising packet contains the RSSI and the device ID.

\begin{dfn} Given three time series data streams, $\Omega_1(\Delta_k)$, $\Omega_2(\Delta_k)$, and $\Omega_3(\Delta_k)$ captured by $\pi_1$, $\pi_2$, and $\pi_3$ during a time window $\Delta_k$, for each device captured by the three sniffers, \emph{queue detection problem} is to determine if the device is in the queue which is starting from $\pi_1$.
\end{dfn}

\subsection{System Design}
\label{SystemDesign}
\Fig{Fig:SystemOverview} shows our system design for queuing behaviour detection which is composed of four layers: \emph{data sources}, \emph{gateways}, \emph{data analytics}, and \emph{applications}. BLE sniffers serve as data sources which capture all types of BLE packets from users' devices and report to the gateway layer. The crowd mobility sensing gateway registers the data types of interests and keeps updating on new data arrivals, where BLE advertising packets are in our interests. Meanwhile, the crowd mobility sensing gateway updates the crowd mobility database when there is new update on received BLE advertising packets. In addition to the crowd mobility database, there are three data analytical components: (1) \emph{data preprocessing}, (2) \emph{feature extraction}, and (3) \emph{queue behavior model training and classification}. The data preprocessing performs data aggregation and low-pass filtering to eliminate noise form the collected RSSIs. The feature extraction identifies the mobility patterns from three aspects: single-device features, cross-device features, and cross-sniffer features. Single-device features are extracted from each individual device's RSSIs. Cross-device features are extracted from multiple user devices' RSSIs based on their mobility similarity. Cross-sniffer features are extracted from observations by multiple sniffers based on the correlation of their observations. Finally, the queue behavior model training and classification conducts off-line training for detecting the status of a device, where the trained classifier makes a binary decision on the device status (e.g., "in-queue" or "not-in-queue" statuses). Finally, the application layer is a knowledge consumer which requires results of queuing detection for improving user experience or services provided by operators.

\section{Queuing Behaviour Data Analytics} \label{Sec:DataAnalytics}
In this section, we first describe the methodology for collecting labeled data and explain the the proposed algorithm for feature extraction. Then, we describe the means for queuing behaviour classification.

\subsection{Labeled Data Collection}
\begin{figure}[!t]
\centering
\begin{tabular}{cc}
\includegraphics[width=0.53\columnwidth]{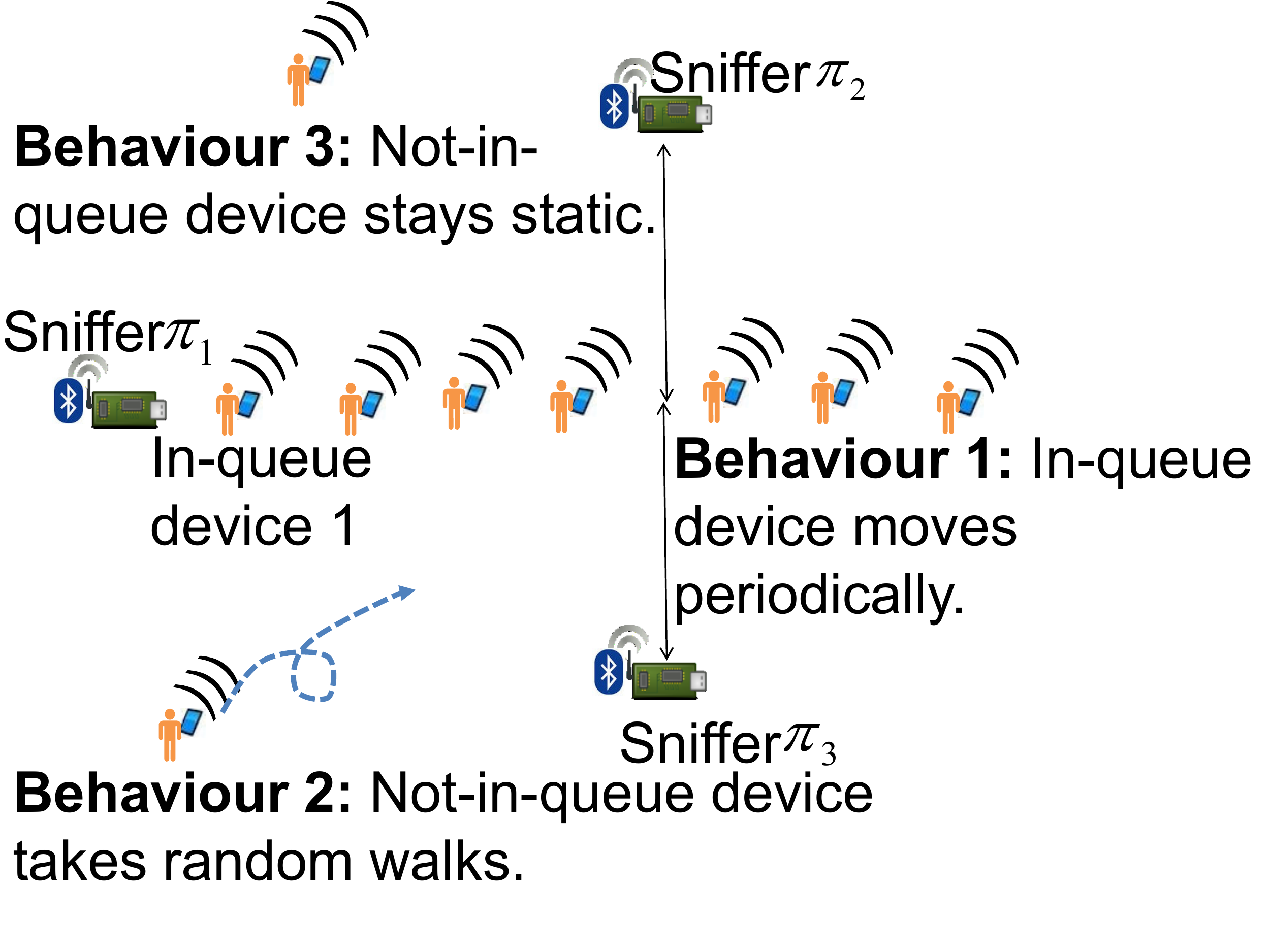}&
\includegraphics[width=0.42\columnwidth]{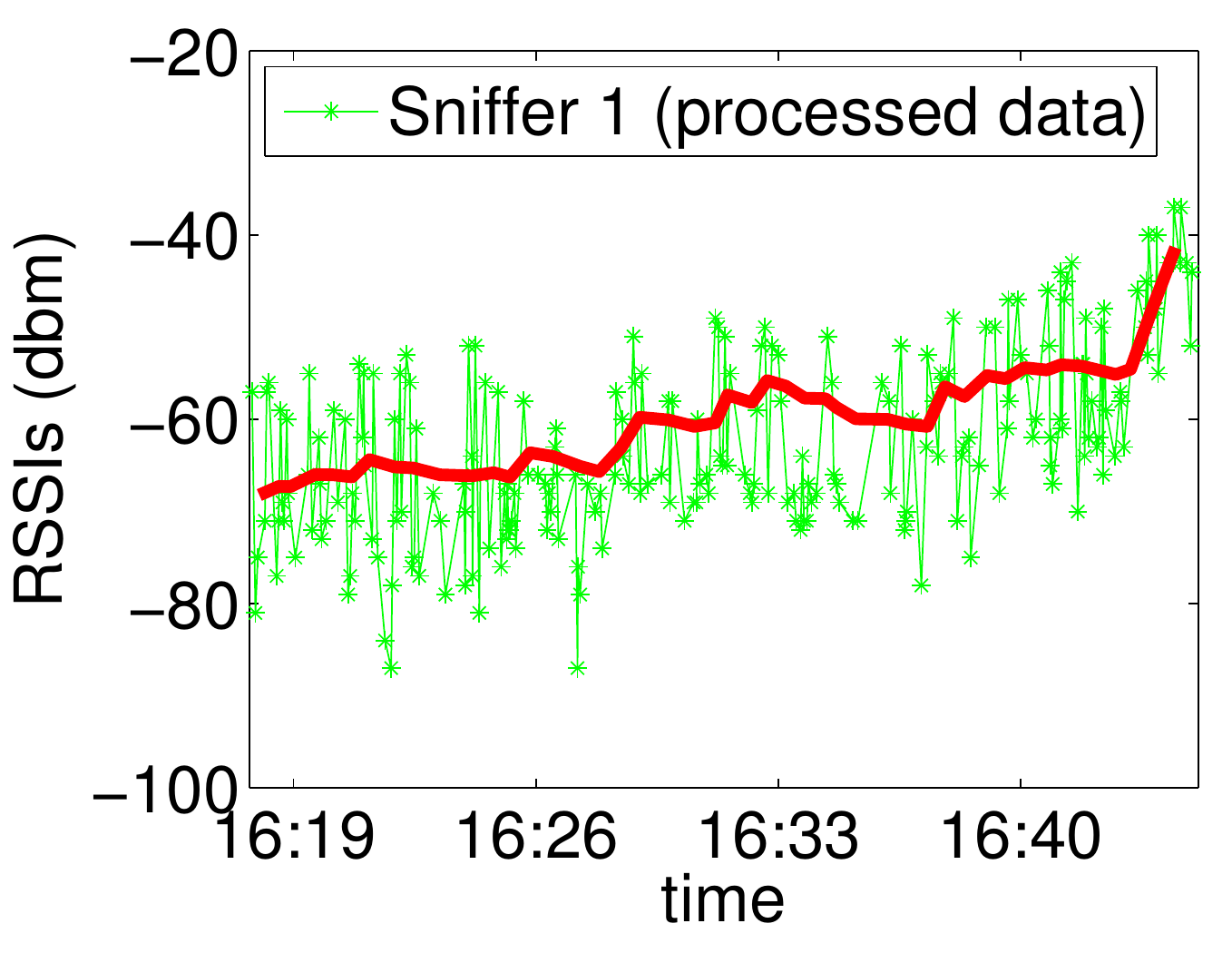} \\
(a) & (b)
\end{tabular}
\caption{(a) Deployment for collecting labelled data. (b) Data preprocessing results.} \label{Fig:DataCollectionAndPreprocessing}
\end{figure}

\Fig{Fig:DataCollectionAndPreprocessing}~(a) is our deployment for collecting labeled data, where human mobility behaviour can be categorized into three types. The first type of mobility behaviour is collected from in-queue devices which have movements periodically. The second type of mobility behaviour is collected from not-in-queue devices which take random walks. The third mobility behaviour is collected from not-in-queue devices which stay static at certain locations. We launch such data collection campaigns in an indoor office environment for collecting labeled data.

\subsection{Data Preprocessing}
Since RSSIs are noisy, we perform data aggregation and low pass filter to clean up noise in the collected raw data. First, in the data aggregation phase, we aggregate RSSI data streams from each device every $\lambda$ seconds. Then, we apply the dynamic exponential smoothing filter (DESF) to the aggregated RSSI data streams. We implement the DESF as follows. The $i$-th output sample is $O_i= \alpha \cdot O_{i-1}+(1-\alpha) \cdot I_{i}$ if $I_{i} < O_{i-1}$. Otherwise, $O_i= (1-\alpha) \cdot O_{i-1}+\alpha \cdot I_{i}$. Here, $I_{i}$ is the $i$-th input sample and $\alpha$ is a predefined parameter. \Fig{Fig:DataCollectionAndPreprocessing}~(b) shows the data before and after data preprocessing, where the green one is raw data and the red one is the processed data after data aggregation and low-pass filtering.

\subsection{Feature Extraction}
We extract three types of features, (1) \emph{single-device features}, (2) \emph{cross-device features}, and (3) \emph{cross-sniffer features}, from preprocessed data. The first type is mobility characteristic extracted from each individual device's RSSIs. The second type is mobility similarity between crowds' devices. The third type is the mobility correlation observed by multiple observers (e.g., sniffers). Our work extracts nine features in total from the three aspects. Below, we first make observations from the collected ground truth data to identify significant features of in-queue devices and then define the mechanisms to extract the features for the off-line training in the next step. The details of the proposed feature extraction mechanisms are explained in the following subsections.

\begin{figure*}[!t]
\centering
\begin{tabular}{ccc}
\includegraphics[width=0.29\linewidth]{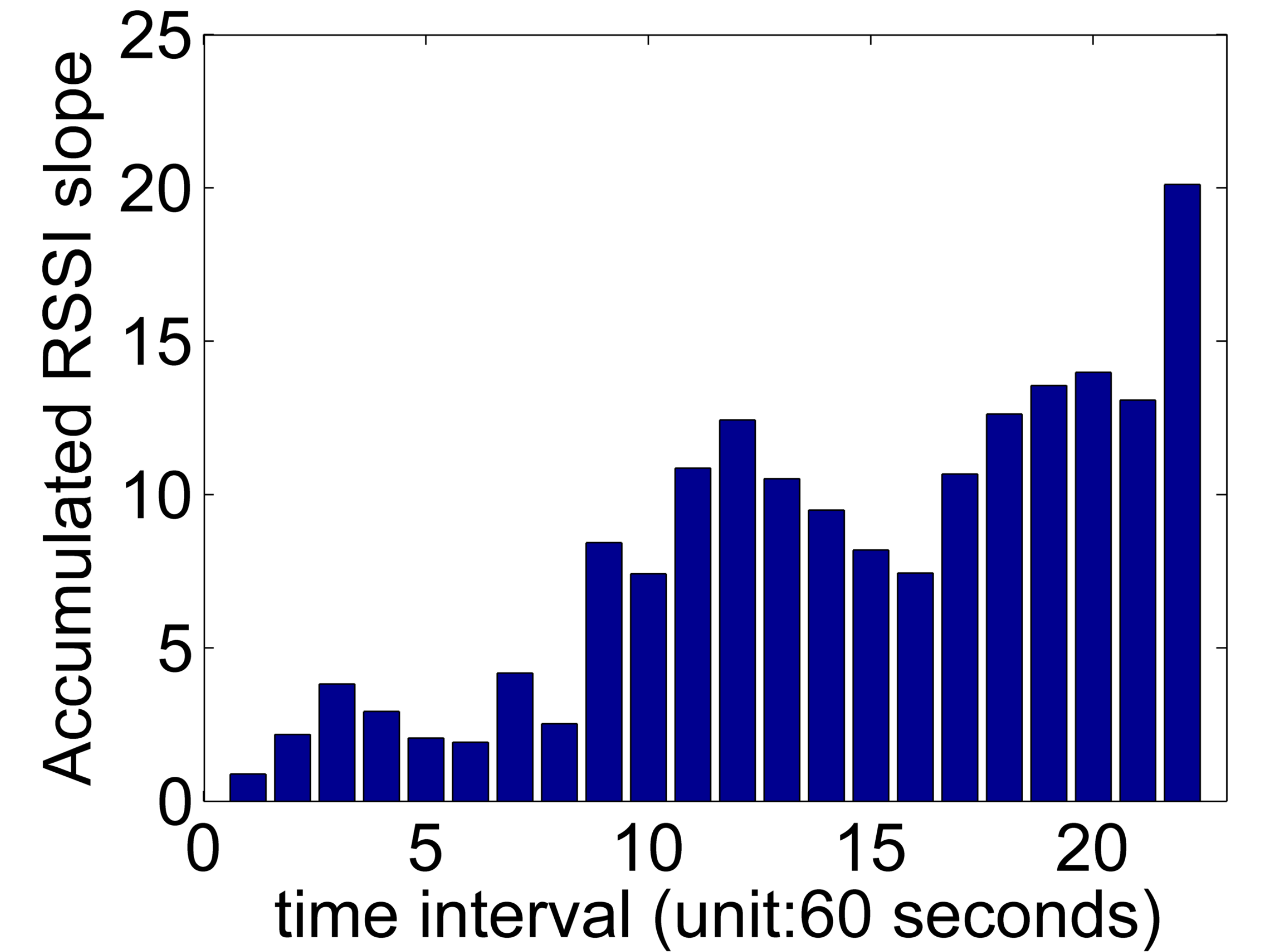} &
\includegraphics[width=0.29\linewidth]{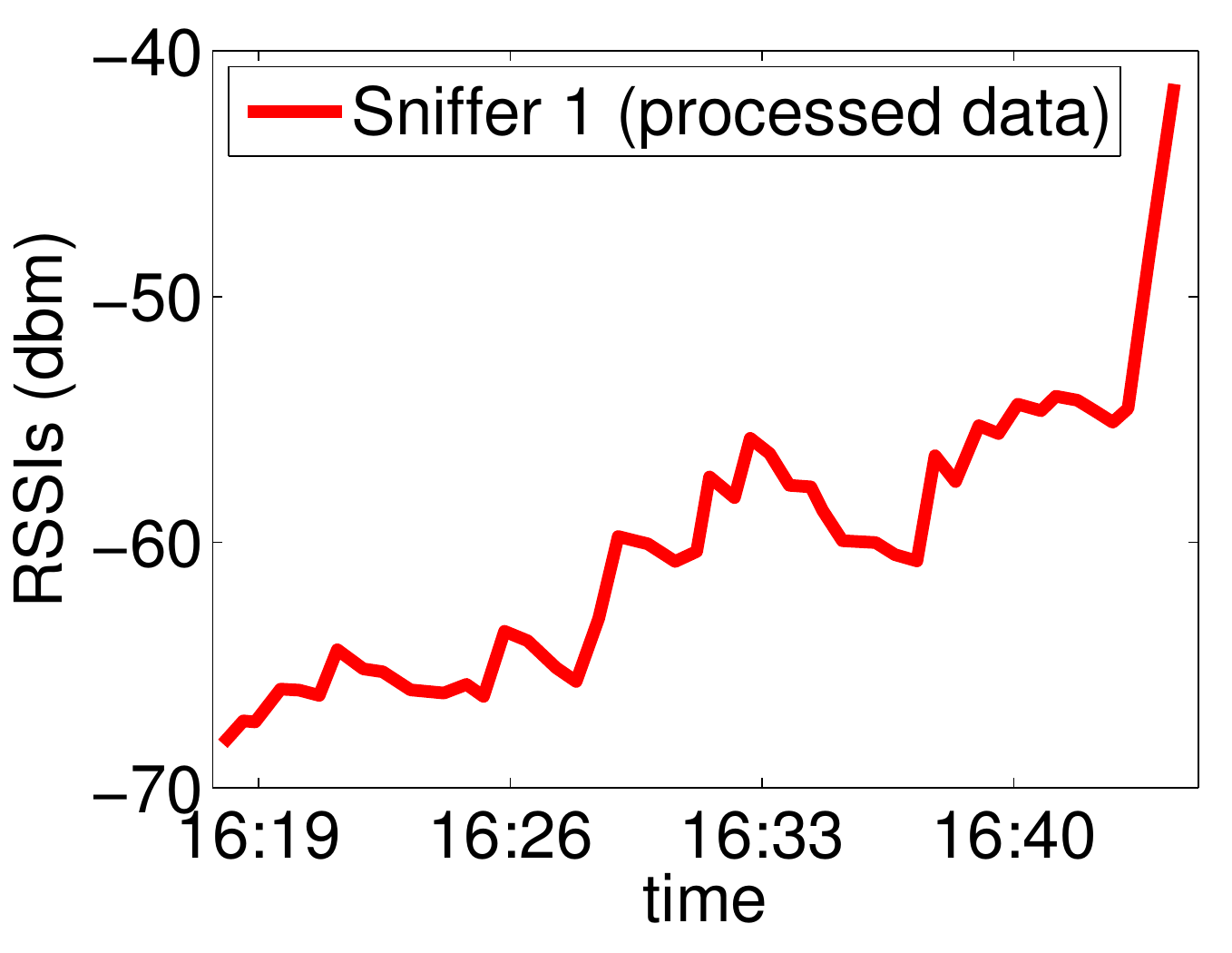} &
\includegraphics[width=0.29\linewidth]{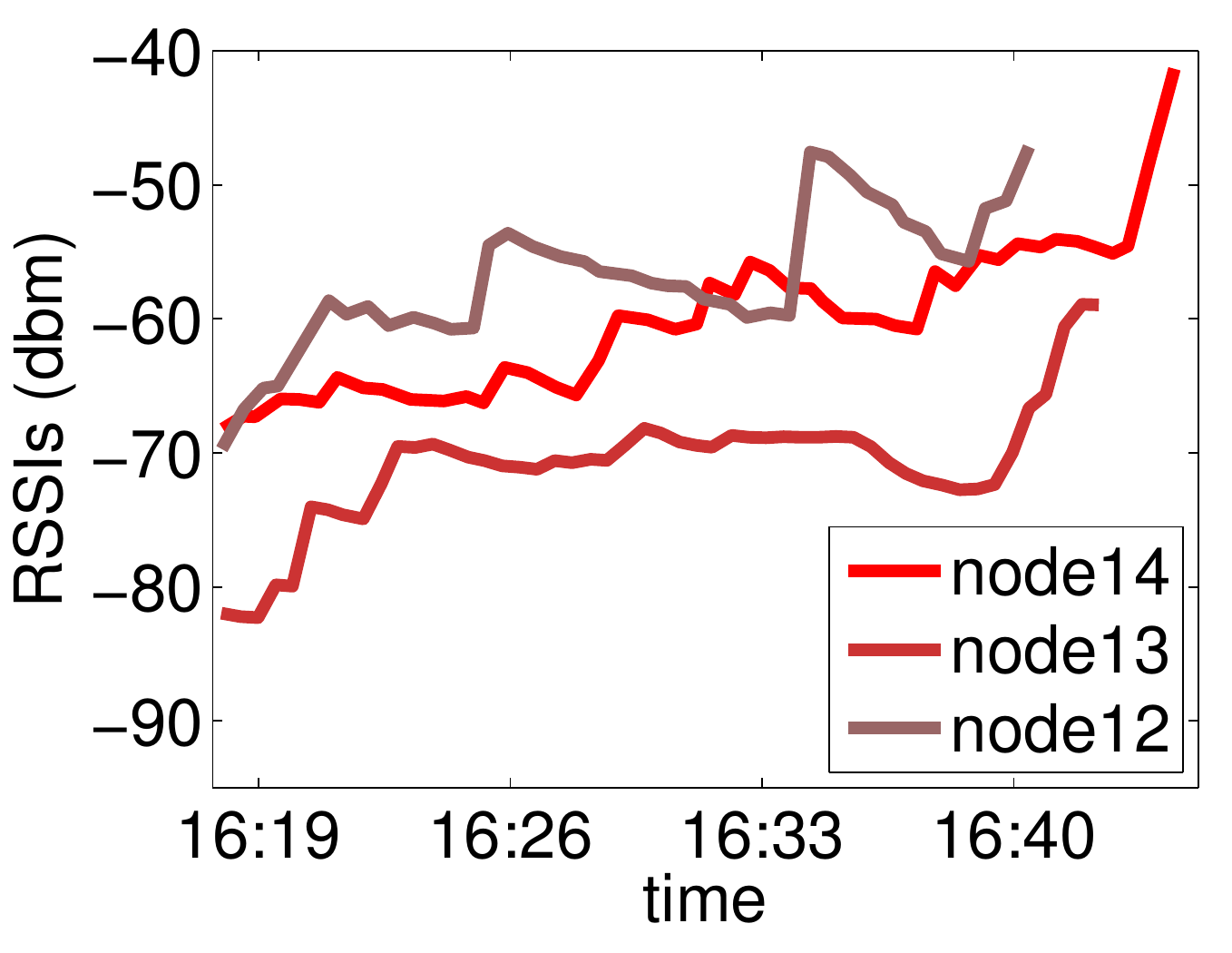} \\
(a) & (b)  & (c) \\
\end{tabular}
\caption{Single-device features: (a) Feature 1: positive accumulated slopes, (b) Feature 2: approaching-counter patterns, and (c) Feature 3: near-counter RSSIs.}
\label{Fig:single-device-features}
\end{figure*}

\subsection*{C.1 Single-device feature extraction}
Let $d_x(\Delta_k)$ denote the captured BLE packets during time window $\Delta_k$ for device $d_x$. For a given time window $\Delta_k=[t_i, t_j)$, for an observed device $d_x$, we define seven feature extraction functions to compute the seven single-device features based on only RSSIs as follows.

\emph{Feature 1: Positive accumulated slopes}: \Fig{Fig:single-device-features}~(a) shows the accumulated RSSI slopes of an in-queue device. As it can be seen, the accumulated RSSI slopes are all positive as the device moves closer to the starting point. Thus, we define the accumulated slope for a given time window $[t_i, t_j)$ as $f_1(d_x, \Omega_1(\Delta_k))=r_j-r_i$, where $r_i$ and $r_j$ are the RSSIs at the timestamps $t_i$ and $t_j$, respectively.

\emph{Feature 2: Approaching-counter patterns}: In \Fig{Fig:single-device-features}~(b), when the in-queue device is already very close to the starting point, the RSSIs dramatically increase even though the in-queue device makes only a little movement towards the starting point. Based on the experimental observation and the signal propagation theory, the function of distance to RSSI changes is not linear. We then define the following binary function to extract this nature feature pattern from RSSIs:
\[
f_2(d_x, \Omega_1(\Delta_k)) =
   \begin{cases}
    1 &\text{if $r_j-r_i > \tau_{f_2}$};\\
    0 &\text{otherwise.}
   \end{cases}
\]
Here, $\tau_{f_2}$ is a pre-defined threshold. When there are dramatic changes on RSSIs during $\Delta_k$, the output value of the binary function is 1 since the device is approaching closer to the starting point. Otherwise, the output of the binary function is 0.

\emph{Feature 3: Near-counter RSSIs}: \Fig{Fig:single-device-features}~(c) shows that the RSSIs changes as these in-queue devices move closer to the starting point. As it can be seen, when devices stay near the staring point, the RSSIs are higher than devices far away from the starting point. We thus define the following binary function to extract this feature based on RSSIs.

\[
f_3(d_x, \Omega_1(\Delta_k)) =
   \begin{cases}
    1 &\text{if $r_t > \tau_{f_3}, \forall t\in \Delta_k$};\\
    0 &\text{otherwise.}
   \end{cases}
\]

Here, $\tau_{f_3}$ is a pre-defined threshold. When all of captured RSSIs during the time window $\Delta_k$, the output value of the binary function is 1 since the device is considered very close to the starting point. Otherwise, the output value of of the binary function is 0.

\emph{Feature 4: RSSI stability observed by $\pi_1$}: The RSSI variances of in-queue devices are smaller compared to the RSSI variances of random walks since in-queue devices make movements slowly along the queue towards $\pi_1$. To track the RSSI changes, we consider $b$ backtracking time windows to extract this feature patterns. For each given time window $\Delta_k$, we backtrack $b$ time windows together with the current observations during $\Delta_k$ to compute RSSI variance as
\[
f_4(d_x, \Omega_1(\Delta_k))=Var(\Omega_1(\Delta_k), \Omega_1(\Delta_{k-1}), \ldots, \Omega_1(\Delta_{k-b})).
\]
Here, $Var(\Omega_1(\Delta_k), \Omega_1(\Delta_{k-1}), \ldots, \Omega_1(\Delta_{k-b}))$ is the variance function of given sets of captured BLE packets' RSSIs during these historical time windows.

\emph{Feature 5: RSSI stability observed by $\pi_2$}: Similarly, we can extract RSSI stability observed by the sniffer $\pi_2$ as:
\[
f_5(d_x, \Omega_2(\Delta_k))=Var(\Omega_2(\Delta_k), \Omega_2(\Delta_{k-1}), \ldots, \Omega_2(\Delta_{k-b})).
\]
Here, the size of the backtracking time windows is the same as in the feature 4.

\emph{Feature 6: RSSI stability observed by $\pi_3$}: Similarly, we can extract RSSI stability observed by the sniffer $\pi_3$ as:
\[
f_6(d_x, \Omega_3(\Delta_k))=Var(\Omega_3(\Delta_k), \Omega_3(\Delta_{k-1}), \ldots, \Omega_3(\Delta_{k-b})).
\]

\emph{Feature 7: Longer stay duration observed by $\pi_1$}: In-queue devices have longer stay durations compared to not-in-queue devices which have wandering random walks. We thus define the following function to extract this feature from packets captured by the sniffer $\pi_1$.
\[
f_7(d_x, \Omega_1(\Delta_k))=t_l-t_f,
\]
where $t_f$ is is the timestamp when the latest advertising packet received by the sniffer $\pi_1$ and $t_f$ is the timestamp when the first advertising packet is received by the sniffer $\pi_1$.

\begin{figure}
\centering
\includegraphics[width=0.8\columnwidth]{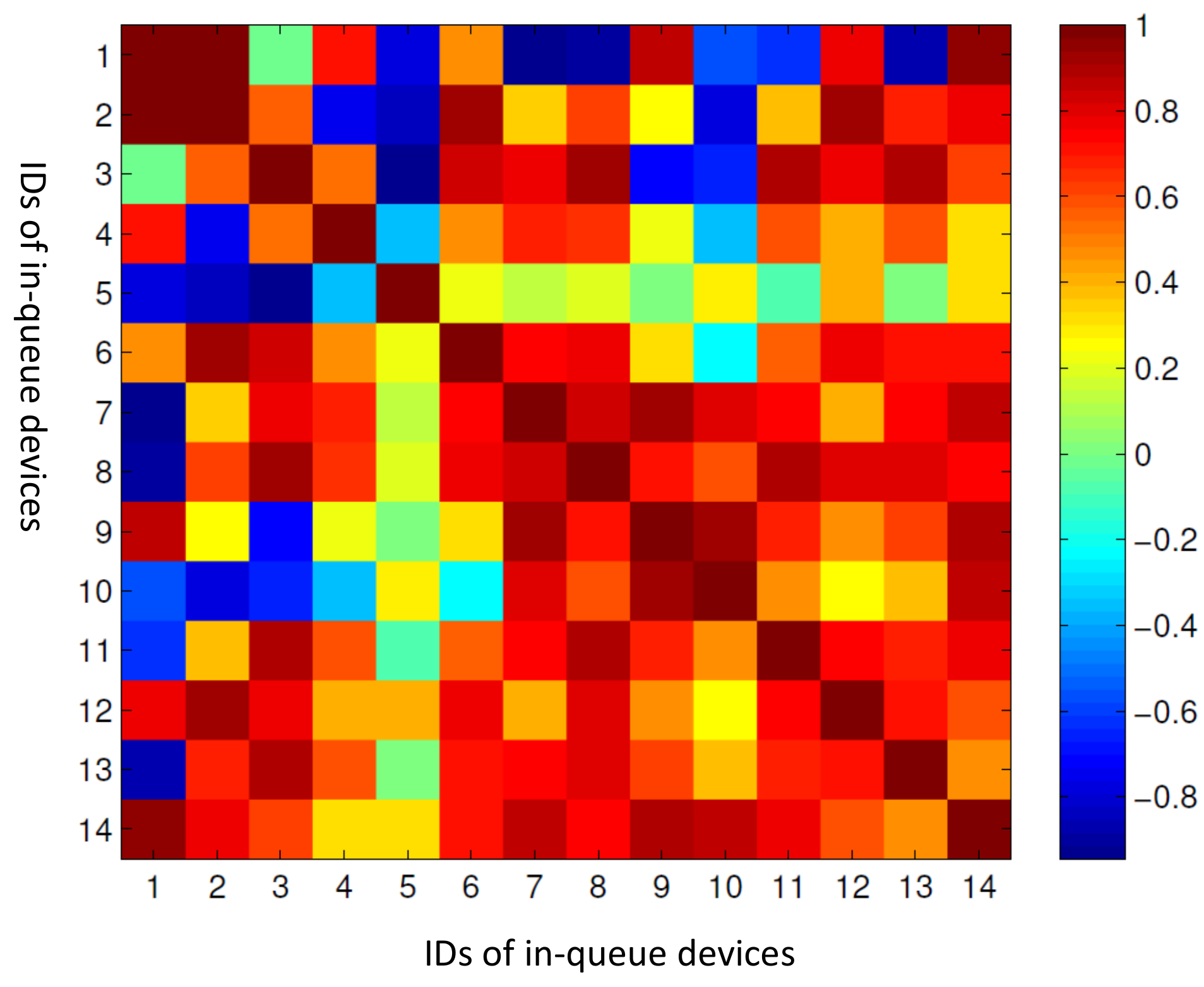}
\caption{Cross-device feature extraction.} \label{Fig:MobilitySmilarity}
\end{figure}

\subsection*{C.2 Cross-device feature extraction} Since queuing behaviour is created by crowds and not by an individual, similar mobility behaviour of crowds creates similar RSSIs patterns to each other. Next, we first identify such kind of cross-device features based on experimental observations and define feature extraction function to speed up computation.

\emph{Feature 8: Mobility similarity}: \Fig{Fig:MobilitySmilarity} shows the visualization of correlation matrix between all pair of in-queue devices' RSSIs captured by the sniffer $\pi_1$. As it can be seen, there are high correlations between RSSIs of in-queue devices when they are queuing up for a certain purpose. Although the sensitivity of devices' antennas are different from each other as shown in \Fig{Fig:single-device-features} (c), their RSSI patterns are similar to each other. Devices far away from the sniffer $\pi_1$ have higher correlation coefficients with others because of sufficient RSSI data samples. However, computing movement similarities between all pairs of devices incurs high computational complexity in the real world since there are many combinations of devices. Therefore, we design the following low-complexity feature extraction function to speed up the computation.
\begin{enumerate}

  \item Sort devices based on the RSSI stability observed by $\pi_1$ (i.e., $f_4(d_x, \Omega_1(\Delta_k))$). The idea is to cross-check those devices with more stable RSSIs first because they are probably in the queue.
  \item Let $\omega_x^{\pi_1}(\Delta_k)$ and $\omega_y^{\pi_1}(\Delta_k)$ denote the BLE packets captured by $\pi_1$ during time window $\Delta_k$ for devices $d_x$ and $d_y$, respectively. The movement similarity function between $\omega_x^{\pi_1}(\Delta_k)$ and $\omega_y^{\pi_1}(\Delta_k)$ is defined as
      \begin{align*}
        S(\omega_x^{\pi_1}(&\Delta_k), \omega_y^{\pi_1}(\Delta_k))\\
        =Cor( &(\omega_x^{\pi_1}(\Delta_k), \omega_x^{\pi_1}(\Delta_{k-1}), \ldots, \omega_x^{\pi_1}(\Delta_{k-b})) \\
              &(\omega_y^{\pi_1}(\Delta_k), \omega_y^{\pi_1}(\Delta_{k-1}), \ldots, \omega_y^{\pi_1}(\Delta_{k-b}))).
      \end{align*}
      Here, $Cor(\cdot,\cdot)$ computes the correlation coefficient between two given sequences of RSSIs in BLE packets $(\omega_x^{\pi_1}(\Delta_k), \omega_x^{\pi_1}(\Delta_{k-1}), \ldots, \omega_x^{\pi_1}(\Delta_{k-b}))$ and BLE packets $(\omega_y^{\pi_1}(\Delta_k), \omega_y^{\pi_1}(\Delta_{k-1}), \ldots, \omega_y^{\pi_1}(\Delta_{k-b}))$. We backtrack $b$ time windows to compute the correlation coefficient between $d_x$ and $d_y$.
  \item For a given $\omega_x^{\pi_1}(\Delta_k)$ and the sorted list in 1), we can compute movement similarity between $d_x$ and devices in the sorted list one by one until there are at least $m$ devices which have correlation coefficients greater than $\tau_{f_8}$. In this case, the $f_8(d_x, \Omega_1(\Delta_k)) = 1$. Otherwise, $f_8(d_x, \Omega_1(\Delta_k)) = 0$. Here, $m$ and $\tau_s$ are predefined thresholds.
\end{enumerate}
Note that the above computation for this feature extraction can be terminated early if we can find $m$ devices which meet the above condition. In this case, we can speed up the computation instead of computing all pairs of combinations.

\begin{figure}
\centering
\includegraphics[width=0.9\columnwidth]{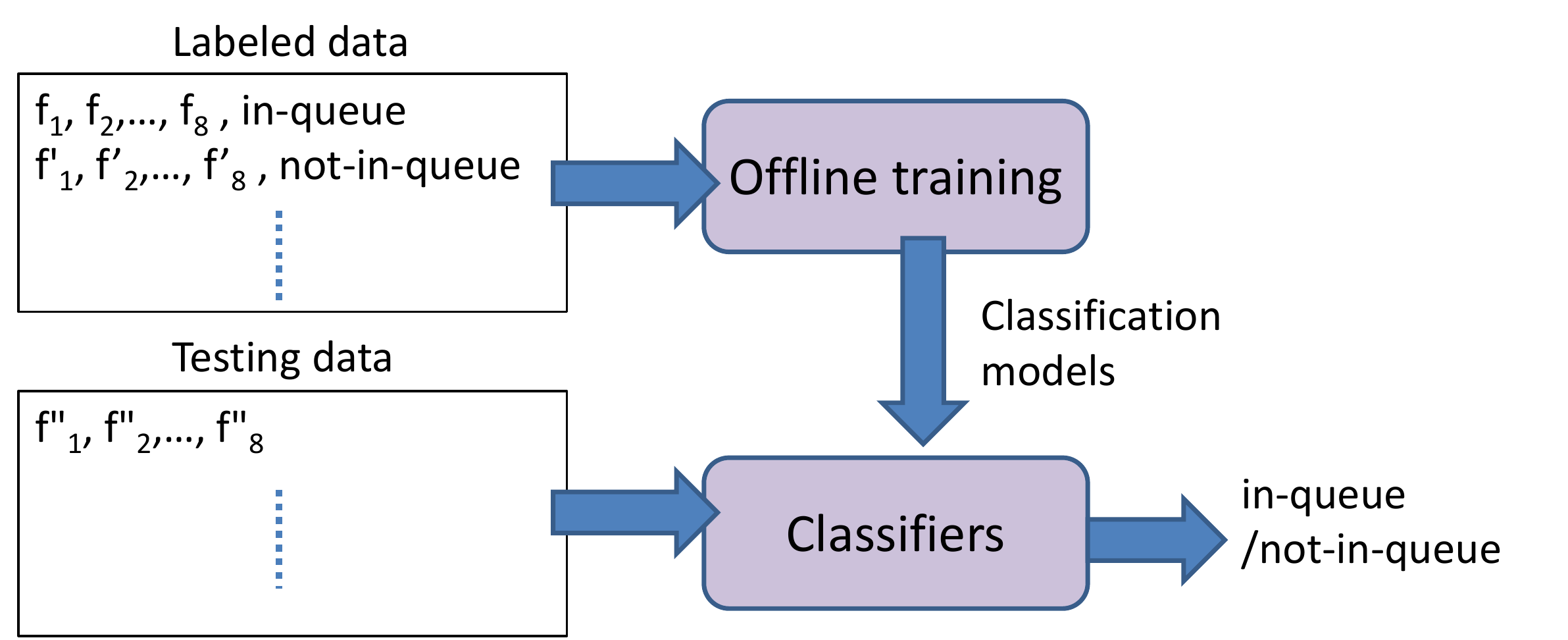}
\caption{The workflow of the queue behavior model training and classification.} \label{Fig:Classification}
\vspace{-0.1cm}
\end{figure}

\begin{figure}
\centering
\includegraphics[width=\columnwidth]{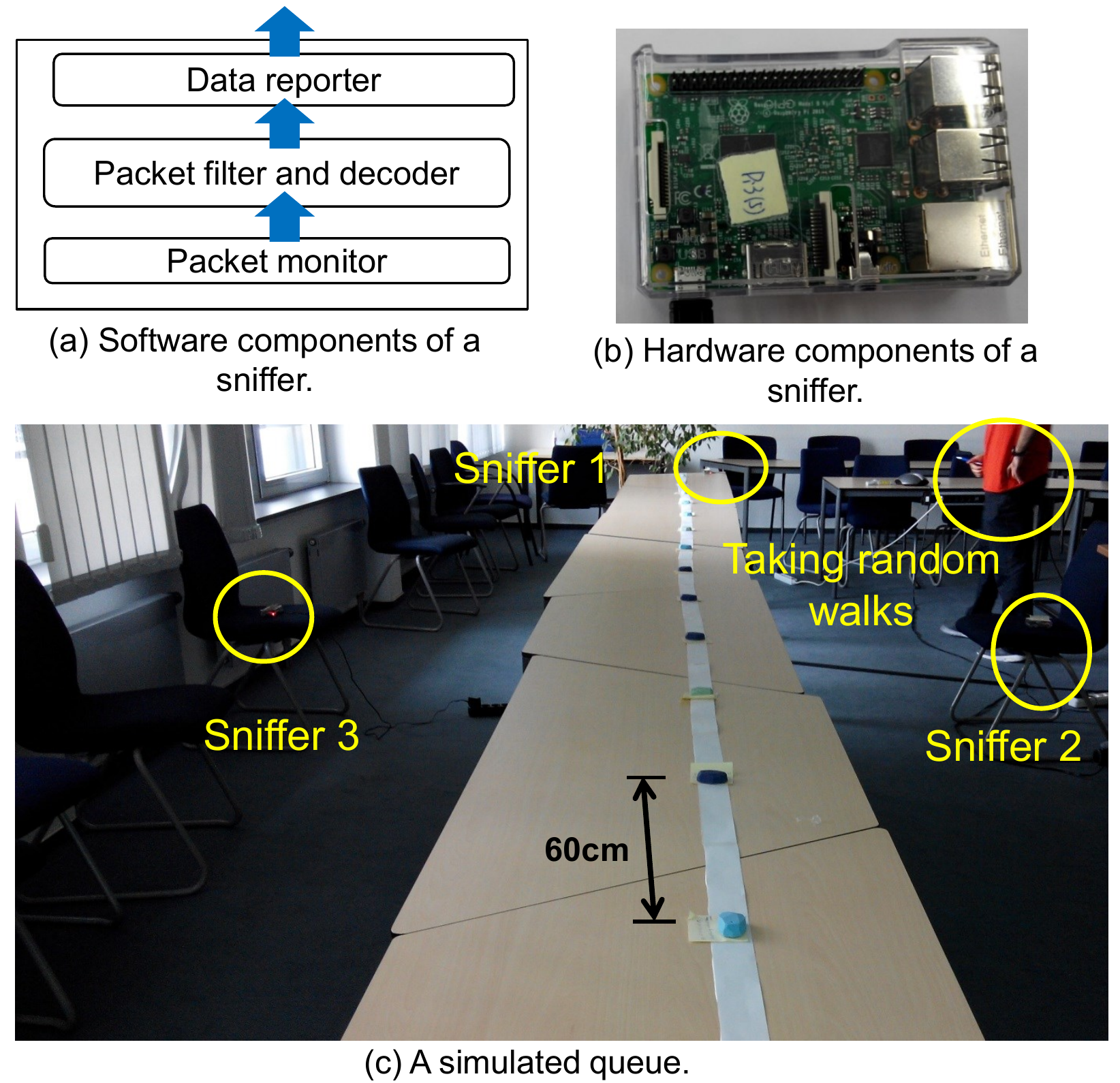}
\caption{Implementation of sniffers and controllable experimental setup with a simulated queue.} \label{Fig:Implementation}
\end{figure}

\subsection*{C.3 Cross-sniffer feature extraction} When an in-queue device makes a sequence of movements along the queue, there is a high correlation between the observations captured by the sniffers on both sides. Note that this feature exploits consensus between sniffers, while cross-device features focus on the mobility similarities between devices. Thus, we consider the cross-sniffer consensus to extract this feature as follows.

\emph{Feature 9: Mobility correlation}: For a device $d_x$, the mobility correlation observed by the sniffer $\pi_2$ and $\pi_3$ during time window $\Delta_k$ is defined as follows.
\begin{align*}
f_8(d_x, \Omega_2(&\Delta_k), \Omega_3(\Delta_k))\\
=Cor(& (\omega_x^{\pi_2}(\Delta_k), \omega_x^{\pi_2}(\Delta_{k-1}), \ldots, \omega_x^{\pi_2}(\Delta_{k-b})), \\
     & (\omega_x^{\pi_3}(\Delta_k), \omega_x^{\pi_3}(\Delta_{k-1}), \ldots, \omega_x^{\pi_3}(\Delta_{k-b}))).
\end{align*}
Here, we backtrack $b$ time windows to compute the correlation coefficient for a device $d_x$'s RSSIs captured by $\pi_2$ and $\pi_3$.

\begin{figure}
\centering
\includegraphics[width=0.75\columnwidth]{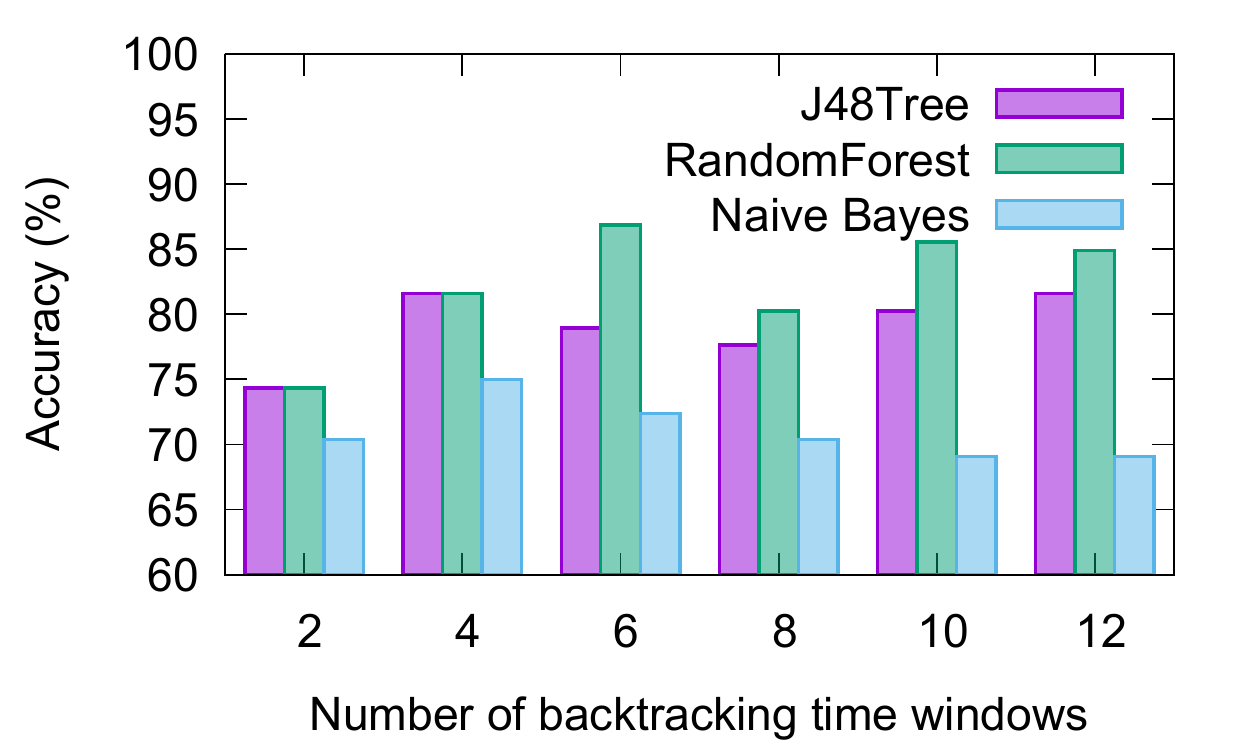}
\caption{Detection accuracy vs. the number of backtracking time windows.} \label{Fig:ExpResults-backtrackingTW}
\end{figure}




\begin{figure}
\centering
\includegraphics[width=0.75\columnwidth]{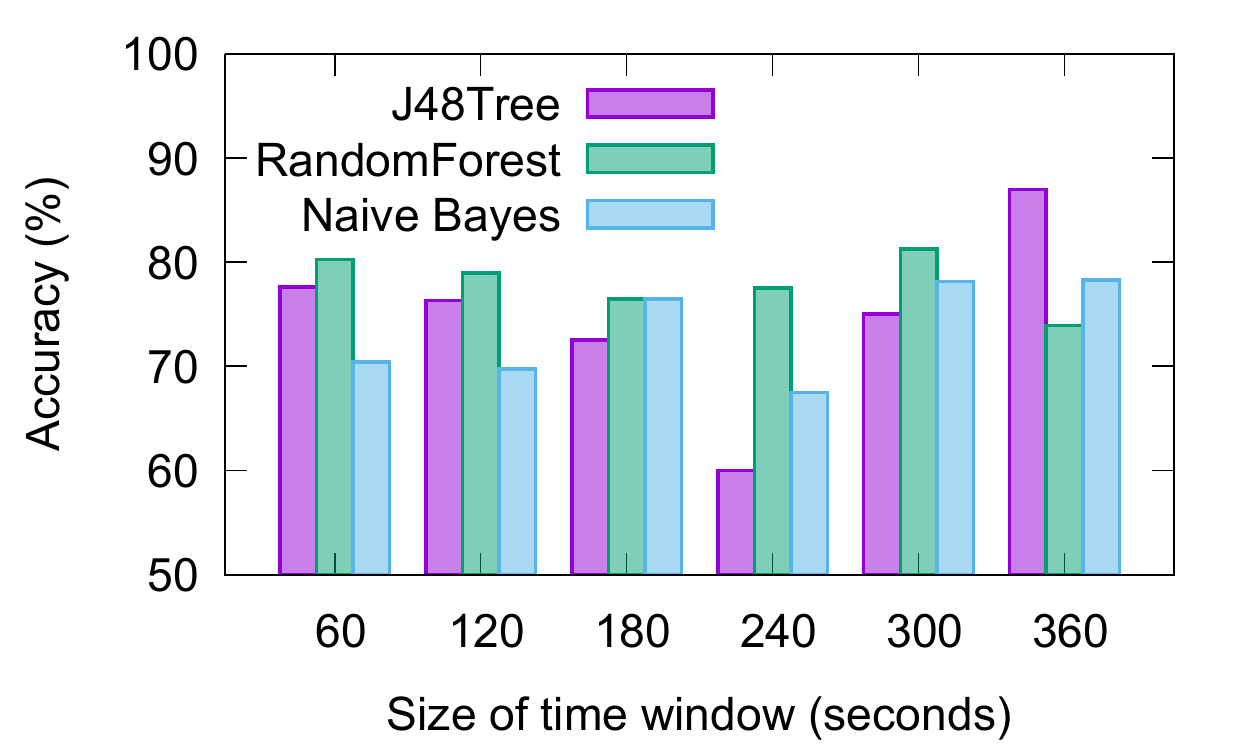}
\caption{Detection accuracy vs. the size of time window.} \label{Fig:ExpResults-TWSize}
\end{figure}



\subsection{Queue Behavior Classification}

\Fig{Fig:Classification} shows the working flow of the designed queue behavior model training and classification. After we extract the above features, these features are used for off-line model training. Here, we consider three existing classifiers: J48Tree, RandomForest, and Naive Bayes since all of these three classifiers work well for numerical and categorical data and incur lower computation overhead which can support real-time detection tasks in the future. After the model training phase, for each piece of the testing data, the classifier determines if the device corresponding to the testing data is in a queue.

\section{Evaluation}\label{Sec:Evaluation}

\subsection{Implementation and Experimental Setup}
We implement a packet sniffing program in Python for Raspberry Pi platforms. We use BLE sensing technology as an example for the prototype of our system. Practically, Wi-Fi sensing technology can be an alternative mechanism to capture human mobility through monitoring probe requests from smartphones. \Fig{Fig:Implementation} (a)-(b) is the software and hardware components of our sniffers. The {\em packet monitor} captures all of types of BLE packets. The {\em packet filter and decoder} parses only BLE advertising packets and discards other types of BLE packets. The {\em data reporter} updates the new packet arrivals to the crowd mobility sensing gateway and then to the crowd mobility database for performing queuing behaviour data analytics described in \Sec{Sec:DataAnalytics}. The designed software components are running as background processes on Raspberry Pi version 3 which has a built-in BLE module. The default parameters in our system are explained as follows. We consider $\lambda=30$ seconds for data aggregation and $\alpha=0.9$ for the low-pass filter in the data preprocessing component. Our system uses a fixed size of time window of 60 seconds. The number of backtracking time windows for extracting RSSI stability (in feature 4, feature 5, and feature 6) is $b=8$ time windows which leads to backtracking for 480 seconds. The pre-defined threshold for extracting the approaching-counter patterns (i.e., feature 2) is $\tau_{f_2}=5$ dbm. The threshold for extracting near-counter RSSIs (i.e., feature 3) is $\tau_{f_3}=-55$ dbm. Finally, we consider $m=3$ and $\tau_{f_8}=0.3$ for extracting nobility similarity (i.e., feature 8).

We conduct two types of experiments: (a) controllable experiments with simulated queuing behaviour and (b) real-world experiments with real human queuing behaviour in a team-building social event. In the controllable experiments, we compare the performance resulted from different classifiers. In the real-world experiments, we compare our signal-based approach against the camera-based face detection approach using OpenCV \cite{OpenCV}.

First, we conduct controllable experiments with a simulated queue to verify our system, where \Fig{Fig:Implementation} (c) shows the experimental setup. Three sniffers are deployed in a big conference room. We deploy 7 BLE beacons on a straight-lined paper to simulate a queue. We pull the straight-lined paper from the starting point of the queue every 120 seconds to simulate human movements along the queue. We use three classifiers, J48Tree, RandomForest, and Naive Bayes, based on Weka \cite{Weka} implementation to evaluate the performance of the queuing behaviour classification. We change the number of backtracking time windows, the size of time window, and the number of sniffers to study how these parameters affect the performance of our system.

Then, moving towards a more realistic environment, the same setup is used in a real-world team building party, where 11 people gather together in an indoor conference room. In the beginning of the team building party, people queue up for taking cakes, biscuits, and drinks, where people casually talk with each other while waiting in the queue. We also deploy a camera in front of the cakes and record videos during the social event.

\begin{figure}
\centering
\includegraphics[width=0.75\columnwidth]{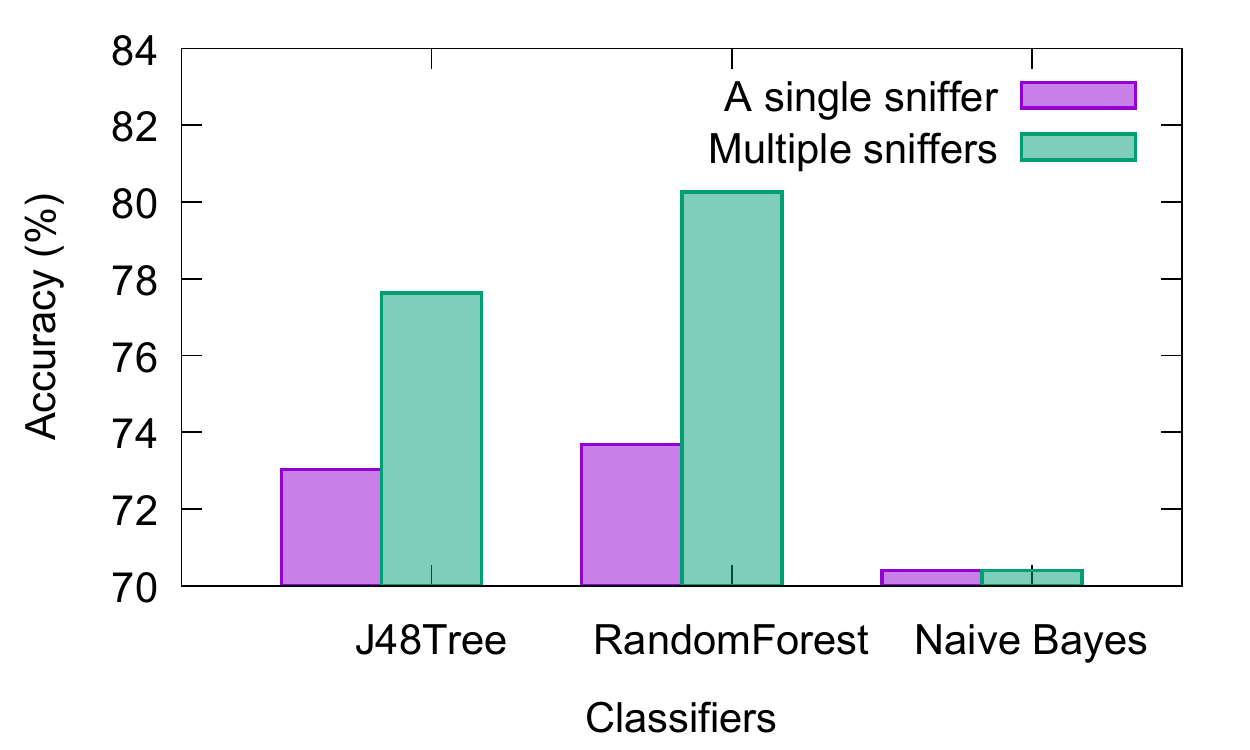}
\caption{Accuracy using different numbers of sniffers.} \label{Fig:changeSnifferNumber}
\end{figure}

\begin{figure*}
\centering
\begin{tabular}{cc}
\includegraphics[width=0.4\linewidth]{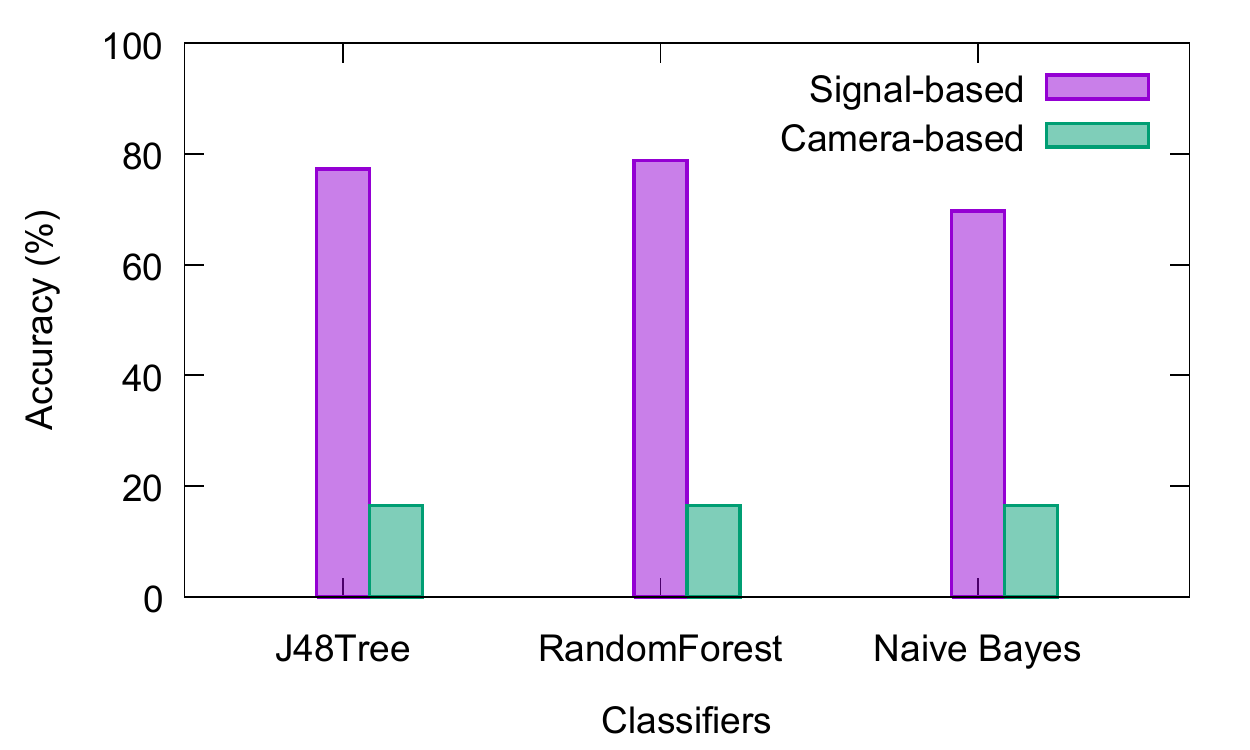} &
\includegraphics[width=0.4\linewidth]{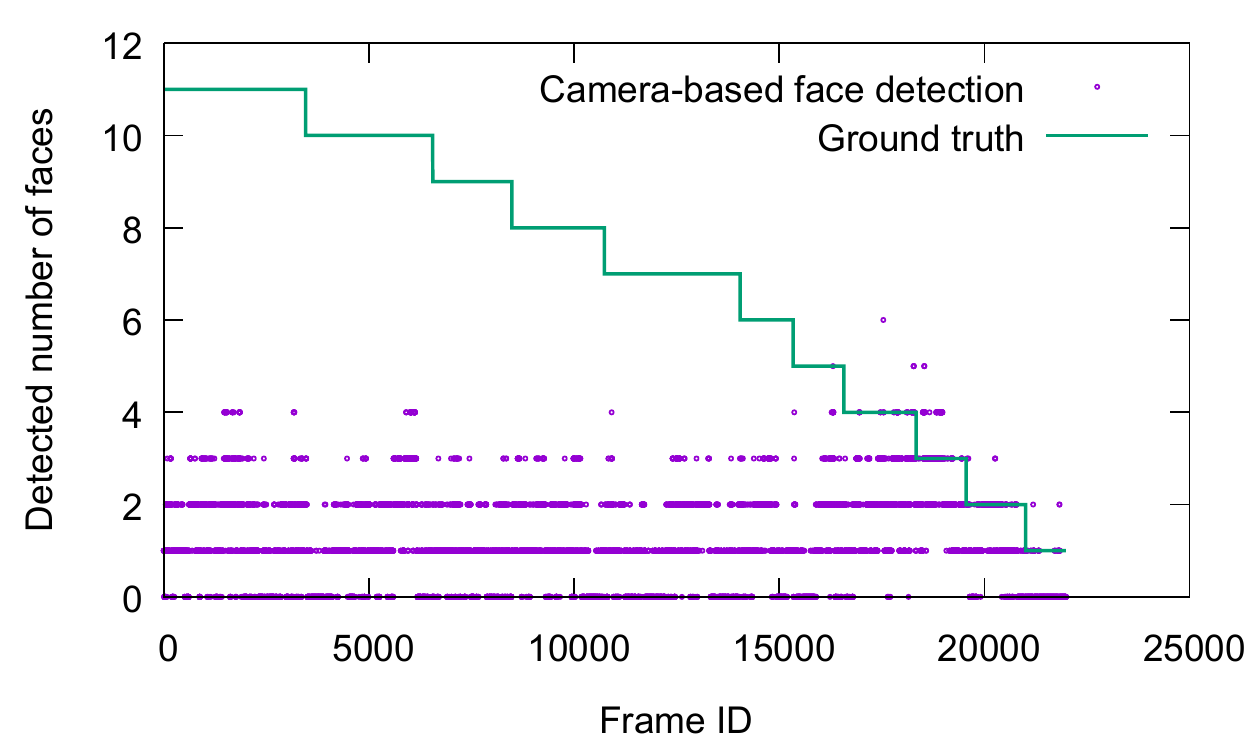} \\
(a) & (b)
\end{tabular}
\caption{Experimental results with a real human queue.} \label{Fig:ExpResults-Camera}
\end{figure*}

\subsection{Experimental Results in a Simulated Queue}

First, we vary the number of backtracking time windows in the feature extraction component from 2 to 12. As shown in \Fig{Fig:ExpResults-backtrackingTW}, RandomForest provides the best accuracy compared to J48Tree and Naive Bayes. We can see that the accuracy resulted from all classifiers increases gradually as backtracking time window increases slightly and becomes stable. This is because a longer backtracking time window involves more mobility information to compute RSSI variances, mobility similarity, and mobility correlation. 

Then, we vary the size of time window from 60 to 360 seconds to extract features. \Fig{Fig:ExpResults-TWSize} shows the accuracy of queuing behaviour detection. We can see that the accuracy decreases as the size of time window increases from 60 to 240 seconds and then increases gradually when the size of time window varies from 240 to 360 seconds. This is because a longer time window considers sufficient RSSIs to compute RSSI variances, mobility similarity, and mobility correlation. However, a too long time window may incur larger RSSI variances because crowds make movements. In this case, RSSI variances cannot provide sufficient insightful information to differentiate in-queue and not-in-queue statuses. 

Next, we compare the detection accuracy using only 1 sniffer and 3 sniffers. \Fig{Fig:changeSnifferNumber} shows the experimental results. As it can be seen, multi-sniffer approaches improve the detection accuracy up to $7\%$. However, the accuracy provided by the Naive Bayes approach cannot be improved since some binary features are considered for model training and Naive Bayes approaches generally work well with numerical data which is used to estimate a distribution over continuous values.

\subsection{Experimental Results with a Real Human Queue}
Finally, we compare our signal-based approach against the camera-based approach in a team-building social event in the real world. During the social event, we record the ground truth by human observations. The recorded video contains 21989 image frames. We use OpenCV face detection libraries to count people in the queue for each frame. \Fig{Fig:ExpResults-Camera} (a) shows the experimental results. We can see that the camera-based face detection approach provides lower accuracy since most of people in the queue are blocked by the people who are closer to the starting point. As shown in \Fig{Fig:ExpResults-Camera} (b), the number of detected faces does not change too much when the queue length becomes shorter. Generally, camera-based approaches are used for crowd mobility monitoring. However, it has limitations to detect queuing behaviour because of blocking visibility issues. We can conclude that signal-based approaches are more suitable for queuing behaviour detection especially for long queues without suffering from blocking issues.

\section{Conclusion}\label{Sec:conclusion}
This paper exploits RSSIs captured by multiple signal sniffers to classify if people are in a queue. We propose three types of feature patterns extracted from each individual's device, cross-device mobility similarity, and cross-sniffer mobility correlation for classification model training. Our approach can be applied to both Wi-Fi and BLE sensing systems. The experimental results indicate that our approach can reach minimum accuracy of $77\%$ and it significantly outperforms the camera-based face detection approach since wireless signals are not blocked by crowds.

\section{Acknowledgment}
\parpic{\includegraphics[width=0.23\linewidth,clip,keepaspectratio]{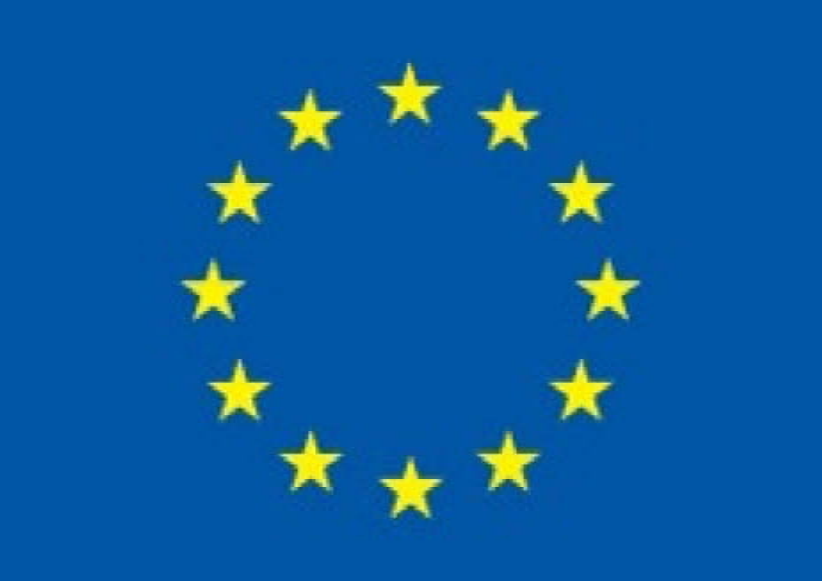}}
\noindent This work has been partially funded by the European
Union's Horizon 2020 research and innovation programme within the
project ``Worldwide Interoperability for SEmantics IoT" under
grant agreement Number 723156.


\bibliographystyle{IEEEtran}                                %


\end{document}